\documentstyle[aps]{revtex}
\begin{document}
\newcommand{\beq}{\begin{equation}}
\newcommand{\eeq}{\end{equation}}
\newcommand{\beqn}{\begin{eqnarray}}
\newcommand{\eeqn}{\end{eqnarray}}
\newcommand{\bmath}{\begin{mathletters}}
\newcommand{\emath}{\end{mathletters}}
\twocolumn[\hsize\textwidth\columnwidth\hsize\csname @twocolumnfalse\endcsname
\title{Overlooked Contribution to the Hall Effect in Ferromagnetic Metals}
\author{J. E. Hirsch}
\address{Department of Physics, University of California, San Diego\\
La Jolla, CA 92093-0319}
 
\date{\today} 
\maketitle 
\begin{abstract} 
It is pointed out that in ferromagnetic metals a contribution to the Hall voltage
arises when a non-zero spin current exists, which is generally the case in the
presence of a charge current. This contribution is independent
of any scattering effects and exists down to zero temperature. The sign of
the resulting Hall coefficient may be either equal or opposite
to the one of the ordinary Hall coefficient depending on the band
filling. This effect seems
to have been left out in previous analyses of the Hall effect in 
ferromagnetic metals.

\end{abstract}
\pacs{}
\vskip2pc]

The Hall coefficient of ferromagnetic metals\cite{hurd,plen}
 is found to be larger than that
of non-magnetic metals and to exhibit a strong 
dependence on the magnetic field ${\bf
B}$.  It is found that the Hall resistivity
\beq
\rho_H=E_y/j_x
\eeq
with $E_y$ the transverse electric field and $j_x$ the longitudinal
current density, can be fitted empirically by the formula
\beq
\rho_{H} = R_{o}B + 4\pi R_{s}M \equiv \rho_H^o +\rho_H^s
\eeq
with B the applied magnetic field and M the magnetization per unit volume. 
$R_{o}$ is the ``ordinary'' Hall coefficient and $R_{s}$ the ``anomalous'' Hall
coefficient.  Various explanations for the origin of $R_{s}$ have been proposed, all
of them involving scattering processes of the conduction electrons together with the
spin-orbit interaction.\cite{hurd,plen}  In some models the carriers are assumed to be
magnetic and the scattering centers non-magnetic,\cite{karp,smit,berg} while in others the situation
is reversed.\cite{kond,mara}

We should point out at the outset that the contribution to the
anomalous Hall effect considered in this paper is not proportional
to the magnetization as given by Eq. (2) but rather to the 
magnetization current. While in some cases the two quantities will
be proportional, this is not necessarily so, and one can even have
situations where a magnetization current exists in the absence of
net magnetization, as discussed later. We will nevertheless use the 
definition Eq. (2) for $R_s$ whenever possible for consistency
with earlier work.

In the theory of Karplus and Luttinger\cite{karp}, the anomalous Hall
effect was explained as arising from interband matrix elements of the
applied electric potential in the presence of spin-orbit coupling in a
perfectly periodic lattice. This theory was criticized by
Smit\cite{smit}, who showed that in fact within the Karplus-Luttinger
treatment a periodic spin-orbit interaction will not give rise to an 
anomalous Hall voltage when all terms are properly taken into
account. Instead, according to Smit, the effect arises from skew 
scattering by perturbations that break the periodicity and lead
to finite resistivity, e.g. impurities and phonons. Later, it was proposed
that in addition to skew scattering a 'side-jump' occurs\cite{berg}
when a magnetic carrier scatters off an impurity, which will also give
a contribution to the anomalous Hall effect. While there has not been
general agreement on whether skew scattering or side jump are
dominant in various cases, and on whether impurity scattering or
phonon scattering dominates, there seems to be a consensus
that the anomalous Hall effect only arises from scattering by
potentials that break the lattice 
periodicity\cite{lutt,hols,nozi,irkh,leri,plen}. Furthermore it is
also generally assumed that the effect is proportional to the magnetization
of the system as given by Eq. (2).

For the case of ferromagnetic transition metals it is generally accepted that the
electrons that give rise to magnetism are itinerant.\cite{wolf}  Here we adopt this point
of view.  The purpose of this paper is to point out that  a
contribution to $R_{s}$ necessarily arises simply from the fact that in general a
spin current exists in ferromagnetic metals
 when a charge current exists. We show that quite generally a 
spin current circulating in a solid will give rise to a transverse Hall field.
  This effect is independent of any scattering
procesess, and seems
to have been omitted in previous discussions of the origin of $R_{s}$ in
ferromagnetic metals.\cite{hurd,plen,karp,smit,berg,kond,mara,lutt,hols,nozi,irkh,leri}
 It gives a contribution to the anomalous
Hall coefficient of the same order of magnitude as the ordinary Hall coefficient.

As is well known, for the ordinary Hall effect there is a 
simple classical explanation. It arises from the Lorentz force that
acts on a moving charge q
\beq
{\bf F}=q\frac{{\bf v}}{c}\times {\bf B}
\eeq
and will appear whenever a charge current circulates in a metal
in the presence of a magnetic field perpendicular to the current
direction. The force Eq. (3) is balanced in steady state by a compensating
electric field
\beq
E_y=\frac{v}{c} B
\eeq
in direction perpendicular to the current and the magnetic field.
Dividing by the current density
\beq
j_x=nqv\eeq
yields the Hall resistivity
\beq
\rho_H=\frac{E_y}{j_x}=\frac{1}{nqc}B
\eeq
where $n$ is the number of electrons in the band if the band
is almost empty, and $q=-e$, with $e$ the magnitude of the electron
charge. If the band is almost full,
Eq. (6) applies with $q=+e$ and $n$ the number of
holes. Of course, for a real metal a quantitative evaluation
of the Hall effect is considerably more complicated
than what is described above. Nevertheless, these simple
considerations capture the essence of the effect.

One may ask whether there is not similarly a simple classical argument that will
predict a transverse Hall field in the presence of a spin current. We show here
 that such an effect is indeed expected. The Hall field will
arise whenever a spin current circulates, with or without charge current 
and with or without net magnetization, in
the presence of a perfectly 
periodic potential, even at zero temperature, just as the ordinary
Hall effect.

Consider an infinite line of equally spaced magnetic moments ${\bf m}$ pointing along
the ${\bf z}$ direction, moving with velocity ${\bf v}$ along the ${\bf x}$
direction, as shown in Figure 1.  An electric field results in the laboratory frame, 
which is identical to
that generated by an infinite line of stationary electric dipoles pointing in the
$(-y)$ direction, given by
\beq
{\bf p} = \gamma {\frac{{\bf v}}{c}} \times {\bf m}
\eeq
with $\gamma = ( 1 - v^2/c^2 ) ^{-1/2}$.
This is seen as follows:  the magnetic field of a magnetic dipole in its rest frame is
\beq
{\bf B} = {\frac{3{\bf n} ({\bf m . n} ) - {\bf m}}{r^{3}}}
\eeq
with ${\bf n}$ a unit vector from the position of ${\bf m}$ to the point where ${\bf
B}$ is observed, and ${\bf r}$ the distance from ${\bf m}$ to the
observation point.  From a Lorentz transformation we find
that there is an electric field in the laboratory frame, given by:
\beq
{\bf E}_{\perp} = - {\frac{\gamma}{c}} {\bf v} \times {\bf B}
\eeq
Here, $(-{\bf v})$ is the velocity of the lab frame with respect to the rest frame of
the moments and $\perp$ indicates directions perpendicular to the velocity.  By
symmetry there is neither electric nor magnetic fields in the direction of ${\bf
v}$.  Eqs. (8) and (9) yield
\beq
{\bf E}_{\perp} = \frac{\gamma}{c} \; \frac{-3({\bf v} \times {\bf n} ) {\bf m.n} +
{\bf v} \times {\bf m}}{r^{3}}
\eeq
On the other hand the electric field from the dipole Eq. (7) is
\beq
{\bf E}_{\perp} = \frac{\gamma}{c} \; \frac{3 {\bf n}_{\perp} ({\bf v} \times {\bf m})
. {\bf n} - {\bf v} \times {\bf m}}{r^{3}}
\eeq
with ${\bf n}_{\perp}$ the projection of ${\bf n}$ in the plane perpendicular to
${\bf v}$.  Although the expressions Eq. (10) and (11) are different for
a single magnetic dipole, when integrated over the infinite line
they yield the same answer. 

Hence we may think of the moving magnetic moments as stationary electric
dipoles, of magnitude given by Eq. (7). 
 In the absence of external potentials there will be no
net transverse force on these dipoles, hence no spontaneous Hall effect.\cite{explain}
However, as we  show next, the periodic lattice potential will exert a transverse
force on these dipoles that tends to deflect them; in steady state, the
transverse force is balanced by accumulation of charge on the edges
under open circuit conditions as in the ordinary Hall effect.

Consider an array of charges $Q$ in a simple cubic lattice of spacing
$a$, as shown in Fig. 2. The electric potential at point $(x,y,z)$ is given
by
\beqn
&V(x,y,z)&=Q\times \nonumber \\
& & \sum_{n_1,n_2,n_3} \frac{1}{[(x-an_1)^2+
(y-an_2)^2+(z-an_3)^2]^{1/2}}
\eeqn
The force in the $y$ direction on an electric dipole of magnitude $p$
pointing in that direction is
\beqn
&F_y(x,y,z)&=p\frac{\partial^2V}{\partial y^2}=
pQ\times \nonumber \\ 
& &\sum_{n_1,n_2,n_3} 
\frac{2(y-an_2)^2-(x-an_1)^2-(z-an_3)^2}{
[(x-an_1)^2+(y-an_2)^2+(z-an_3)^2]^{5/2}} .
\eeqn
The sign of the force depends on the location of the dipole.
In particular, it is negative on $(010)$ planes that go 
through lattice points and positive  on $(010)$ planes midway between
lattice points, as indicated in Figure 2.

Assume that the magnetic moments propagate in rectilinear motion
along the $x$ direction with constant velocity. It is then appropiate
to average the force Eq. (13) over different x-positions in the
unit cell, yielding
\beqn
&\bar{F}_y(y,z)&=\frac{1}{a} \int_0^a dx F_y(x,y,z)=
\frac{pQ}{a}\times \nonumber \\
& &\sum_{n_1,n_2,n_3} [[2(y-an_2)^2-(z-an_3)^2]
G_1-G_2]
\eeqn
with
\bmath
\beq
G_1=F_1(\omega_2)-F_1(\omega_1)
\eeq
\beq
G_2=F_2(\omega_2)-F_2(\omega_1)
\eeq
\beq
\omega_1=-an_1
\eeq
\beq
\omega_2=a(\frac{1}{2}-n_1)
\eeq
\beq
F_1(\omega)=\frac{2\omega(2\omega^2+3r^2)}
{3r^4(w^2+r^2)^{3/2}}
\eeq
\beq
F_2(\omega)=\frac{2\omega^3}
{3r^2(w^2+r^2)^{3/2}}
\eeq
\beq
r^2=(y-an_2)^2+(z-an_3)^2
\eeq
\emath
Eq. (14), with $p$ given by eq. (7), is the transverse force
acting on a magnetic moment that propagates with uniform 
velocity along a straight path parallel to one of the principal
axis of the cubic lattice. We find the remarkable result\cite{find}
\beqn
&<\bar{F}_y>& \equiv 
\frac{1}{2}[\bar{F}_y (r cos \theta ,r sin \theta )+\nonumber \\
& &\bar{F}_y(r cos (\pi /2- \theta ),rsin( \pi /2- \theta ))]=
\nonumber \\
& &\frac{pQ}{a^3}\frac{2 \pi }{3}
\eeqn
$independent$ $of$ $r$ $and$ $\theta$.

Eq. (16) is a central result of this paper. It says that the
transverse force acting on a magnetic carrier, when averaged over a
rectilinear trajectory and its mirror image across the $(0,1,1)$ plane,
$is$ $the$ $same$ $for$ $all$  $trajectories$. Thus the transverse electric field
from Eq. (16) for carriers of charge $q$ is
\beq
E_y=\frac{2\pi}{3}\frac{Q}{qc}\frac{m|v|}{a^3}
\eeq
If there are $\nu$ carriers per atom, all polarized in the same
direction, the magnetization per unit volume is
\beq
M=\frac{\nu m}{a^3}
\eeq
and the ionic charge is $Q=\nu e$ for charge neutrality, so that
\beq
E_y=\frac{2\pi}{3}\frac{|v|}{c} M.
\eeq
For the case of a nearly empty band with spin-polarized electrons, the
charge current is
\beq
j_x=-n_\uparrow ev
\eeq
so that the spontaneous Hall coefficient defined by Eq. (2) is simply
\beq
R_s=-\frac{1}{6n_\uparrow ec}=\frac{R_o}{6}
\eeq

Consider now a general situation in a solid.  The transverse field from the effect
discussed here is obtained by averaging Eq. (19) over all charge carriers.
The particle number current for electrons of spin $\sigma = \uparrow ,
\downarrow$ is:
\beq
{\bf j}_{\sigma} = \sum_{\nu} \int \frac{d^{3}k}{(2 \pi )^{3}} \; g
({\varepsilon^{\nu}_{\sigma}} (k) ) {{\bf v}^{\nu}_{\sigma}} (k)
\eeq
where $g$ is the electron distribution function, $\nu$ labels the bands, and the
velocity is given by 
\beq
{\bf v}^{\nu}_{\sigma} (k) = \frac{1}{\hbar} \; \frac{d {\varepsilon_
{\sigma}} ^{v} (k)}{d{\bf k}}
\eeq
with $\varepsilon^{\nu}_{\sigma} (k)$ the band energy. The charge
current is given by
\beq
j_{ch} = (-e) ({\bf j} _\uparrow + {\bf j}_\downarrow )
\eeq
and the spin current, or more precisely magnetic moment current, by 
\beq
{\bf j}_{spin} = - \mu_B ({\bf j}_\uparrow - {\bf j}_\downarrow)
\eeq
with $\mu_B$ the Bohr magneton.
The Hall field originating from the effect under
discussion here is then
\beq
E_{y} = \frac{2\pi}{3} \frac{\gamma}{c} j_{spin}
\eeq
so that the contribution to the Hall resistivity due to this
effect is
\beq
\rho_{H}^s = \frac{2\pi}{3} \frac{\gamma}{c} \; \frac{j_{spin}}{j_{ch}}
\eeq
and the anomalous Hall coefficient is given by
\beq
R_s=\frac{1}{6}\frac{\gamma}{c}\frac{j_{spin}}{j_{ch}M}
\eeq
with the magnetization $M$ given by
\beq
M=-\mu_B \sum_{\nu} \int \frac{d^{3}k}{(2 \pi )^{3}} \; 
[f({\varepsilon^{\nu}_{\uparrow}} (k) ) -
f({\varepsilon^{\nu}_{\downarrow}} (k) )]
\eeq
with $f$ the Fermi function.

Within semiclassical transport theory and the relaxation time approximation we have
for the current in the presence of an applied longitudinal electric field E:
\beq
{\bf j_\sigma}  =
 (-e) {\sum_{\nu}} \left[ \int \frac{d^{3}k}{(2 \pi)^{3}} \;
{\tau_\nu} ({\bf k}) \left( - {\frac{df}{d \varepsilon}} \right) _{\varepsilon =
{\varepsilon_{\sigma}^{\nu}} (k)} {{\bf v}^{\nu}_{\sigma}} (k)
{{\bf v}^{\nu}_{\sigma}} (k) \right] {\bf E} 
\eeq
with $\tau$ the collision time.  Assuming $\tau$ depends on momentum only through the
band energy and temperatures much smaller than the Fermi energy Eq. (30) can be
rewritten as
\beq
{\bf j}_\sigma = (-e) \left[ {\sum_{\nu}} \tau_{\nu} (\varepsilon_{F}) \int
\frac{d^{3}k}{(2 \pi)^{3}} \; \frac{1}{\hbar ^{2}} \;
\frac{d^{2} {\varepsilon^{\nu}_{\sigma}}}{d{\bf k} d {\bf k}} \; f
(\varepsilon_{\sigma} (k)) \right] {\bf E}
\eeq
Consider in  particular the case of a nearly empty band with isotropic
Fermi surface. We have 
\bmath 
\beq
\frac{1}{\hbar ^{2}} \;
\frac{d^{2} {\varepsilon^{\nu}_{\sigma}}}{d{\bf k} d {\bf k}}
\equiv \frac{1}{m_\sigma}
\eeq
\beq
j_\sigma=\frac{\tau_\sigma n_\sigma}{m_\sigma}eE
\eeq
\beq
M=-\mu_B (n_\uparrow - n_\downarrow)
\eeq
\emath
so that 
\beq
R_s=-\frac{\gamma}{6ec}\frac{
\frac{n_\uparrow \tau_\uparrow}{m_\uparrow}-
\frac{n_\downarrow \tau_\downarrow}{m_\downarrow}}{
(\frac{n_\uparrow \tau_\uparrow}{m_\uparrow}+
\frac{n_\downarrow \tau_\downarrow}{m_\downarrow})
(n_\uparrow - n_\downarrow)}
\eeq
For the case where $m_\uparrow=m_\downarrow$
and $\tau_\uparrow = \tau_\downarrow$ the anomalous
Hall coefficient reduces to Eq. (21). However
because majority and minority spin Fermi surfaces will
be different, the relaxation times could be very different.
Furthermore, when there is spin polarization the
 effective masses of spin up and down electrons
could be very different due to interaction effects.
  If the effective masses
for up  and down electrons are very different Eq. (33) leads to
a  Hall coefficient $R_s$ that can be substantially 
larger than the ordinary one.

Even without assuming different effective masses for up
and down electrons the Hall resistivity from this effect
will in general be different than Eq. (21). Consider again a
single band, and a
Stoner-like description of magnetism.\cite{wolf}  In the magnetic state 
the band energies
are shifted by the exchange splitting $\Delta$:
\bmath
\beq
\varepsilon_{\uparrow} (k) = \varepsilon (k) - \frac{\Delta}{2}
\eeq
\beq
\varepsilon_{\downarrow} (k) = \varepsilon (k) + \frac{\Delta}{2},
\eeq
\emath
the magnetization is given by eq. (29) (for a single band)
and for small $\Delta$ we have
\beq
M = - \mu_{B} g (\varepsilon_{F}) \Delta
\eeq
with $g(\varepsilon_{F})$ the density of states at the Fermi energy.  The spin current
Eq. (25) is
\beq
{\bf j}_{spin} = e\tau(\epsilon_F) \mu_B \Delta \left[ \int \frac{d^{3}k}{(2 \pi ) ^{3}} \;
\frac{1}{\hbar ^{2}} \; \frac{d^{2} \varepsilon (k)}{d {\bf k} d {\bf k}} \left( -
\frac{df}{d \varepsilon} \right) \right] {\bf E}
\eeq
or
\beq
{\bf j}_{spin} = -e\tau(\epsilon_F)
M \langle \frac{1}{\hbar^{2}} \; \frac{d^{2} \varepsilon (k)}{d {\bf k} d {\bf
k}} \rangle {_{F.S.}} {\bf E}
\eeq
where $M$ is the magnetization.
The quantity in brackets is the effective mass tensor
\beq
{\bf M}^{-1} (k) = \frac{1}{\hbar^{2}} \frac{d^{2} \varepsilon (k)}{d {\bf k} d {\bf
k}} \eeq
and it is averaged over the Fermi surface.  In contrast the charge current is given by
\beq
{j_{charge}} = e^2\tau(\epsilon_F)
 {\left[ {\int} {\frac{d^{3}k}{(2 \pi)^{3}}} {\frac{1}{\hbar^{2}}}
{\frac{d^{2}\varepsilon (k)}{d {\bf k} d {\bf k}}} {f(\varepsilon(k))} \right]} {\bf
E}
\eeq
so that it involves the effective mass tensor integrated over the occupied (or empty)
states:
\beq
{\langle {\bf M}^{-1} \rangle _{occ}} = {\frac{\int {\frac{d^{3}k}{(2\pi)^{3}}}
{\frac{1}{\hbar^{2}}} {\frac{d^{2}\varepsilon}{d{\bf k}d{\bf k}}}
{f(\varepsilon(k))}}{\int {\frac{d^{3}k}{(2\pi)^{3}}} {f(\varepsilon(k))}}}
\eeq
Eq. (27) then yields
\beq
{\rho_{H}} = {-\frac{\gamma}{6nec}} {\frac{\langle {\bf M}^{-1}_{ii}\rangle
_{F.S.}}{\langle {\bf M}^{-1}_{ii}\rangle _{occ}}} 4 \pi  M
\eeq
where the number of carriers $n$ is given by the denominator of Eq. (40).  We
have assumed that the applied electric field is along a principal axis of the
effective mass tensor labeled by $i$.

Eq. (41) shows that this contribution to the Hall voltage is proportional to the
magnetization, as Eq. (2) indicates.  The magnitude depends on the detail of the band
structure through the averages of the effective mass tensors given in Eq. (41). 
Although for an almost empty (or almost full) band the two averages in Eq. (41) will
be very similar, in other cases they could be very different.

It is important to emphasize that the anomalous Hall effect discussed here
originates in the spin current and not in the magnetization. Consider for 
example the situations depicted in Figs. 3(a) and (b). In both cases the 
carriers in the spin-up band, that carries the current, are electrons,
 since it is less
than 1/2-full, so that the ordinary Hall coefficient $R_o<0$. However,
in (a) the spin-down band is empty while in (b) it is full. Hence, in (a)
the current carriers have magnetic moment parallel to the overall magnetization
and the effect discussed in this paper predicts an anomalous Hall coefficient
of the same sign as the ordinary one, $R_s<0$. Instead in (b), the 
current carriers have magnetic moment opposite to the overall magnetization,
hence the anomalous Hall coefficient $R_s>0$, opposite in sign to the
ordinary one.

Consider then the Hall voltages for a single band that undergoes full
spin polarization. For simplicity we assume the band is symmetric around its center, as
occurs for example for a tight binding band in a bipartite lattice structure.
At onset of spin polarization the ordinary and anomalous Hall coefficient will
have the same sign, negative for $n<1$ (less than half-filled band)
and positive for $n>1$. For full spin polarization instead, the
ordinary and anomalous Hall coefficients as function of the total
band filling $n=n_\uparrow+
n_\downarrow$ (i.e. the band filling of the unpolarized band) 
have the behavior indicated in Figure 4. In particular, they have
opposite sign for band fillings between one quarter and three quarters.
Note also that $R_s$ is discontinuous at half filling because the spin
current switches sign at that point.

In the presence of several bands the analysis will become more complicated, but it is
clear that a detailed analysis can provide useful information on the band structure
of the metal.  It should also be kept in mind that contributions to the
ordinary Hall coefficient will arise  both from current carriers in
bands involved in the magnetism
as well as those in non-magnetic bands.
Temperature dependence of the effect discussed here
will arise both from the temperature dependence
of the exchange splitting, the relaxation time $\tau (k)$, and possibly
the effective masses.  It is clear that
the temperature dependence of spin and charge currents could be very different, as
the relaxation time will involve different phonons for the majority and minority spin
bands.  Thus the Hall voltage Eq. (26) may exhibit strong temperature dependence, as
observed experimentally.  It is possible that the effect discussed here is an
important contribution to the anomalous Hall effect observed in ferromagnetic metals.

In summary, we have shown here that a Hall field will arise when a spin current
flows in a solid.  This effect should contribute to the
anomalous Hall effect of ferromagnetic metals because flow of a charge
current in the presence of nonzero magnetization will generally 
(although not always) be accompanied by flow of a spin current.
Although in the simplest model this contribution to the Hall effect is
of similar magnitude to that of the ordinary effect,  in
general the contribution will be different and could be substantially larger.
The detailed temperature dependence in various cases should be investigated.

One could also have a situation where a spin current flows in the absence
of magnetization. In a hypothetical half-metallic antiferromagnet\cite{groo},
flow of charge current is accompanied by flow of spin current in the absence
of magnetization, and a transverse Hall field will result. In the model of
spin-split metals discussed in Ref.17  , a spin current flows in the absence of
both magnetization and charge current, and a transverse Hall field should
also exist. Other examples are discussed in Ref. 18, in connection with
manganites, and in Ref. 19 for superfluid $^3He$. Finally, a
 pure spin current is
also predicted to occur in the geometry of the 'spin Hall effect' discussed 
in Ref. 20, where 
opposite edges of a conducting sample are connected through a transverse
conducting strip.

\acknowledgements
The author is grateful to Harry Suhl for helpful discussions.

\newpage

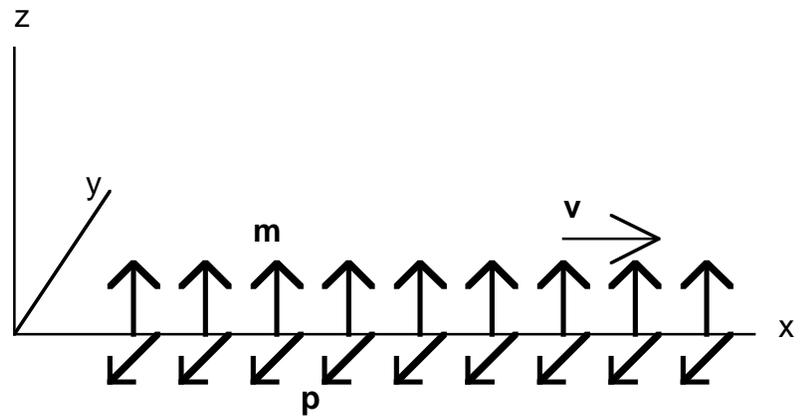
\begin{figure}
\caption {Line of magnetic moments pointing in the z direction
moving along the x direction with 
velocity v. The electric field generated in the laboratory frame
is the same as that obtained from a line of equally spaced 
stationary electric dipole
moments pointing along the (-y) direction, of magnitude given
by Eq. (7). }
\label{Fig. 1}
\end{figure}

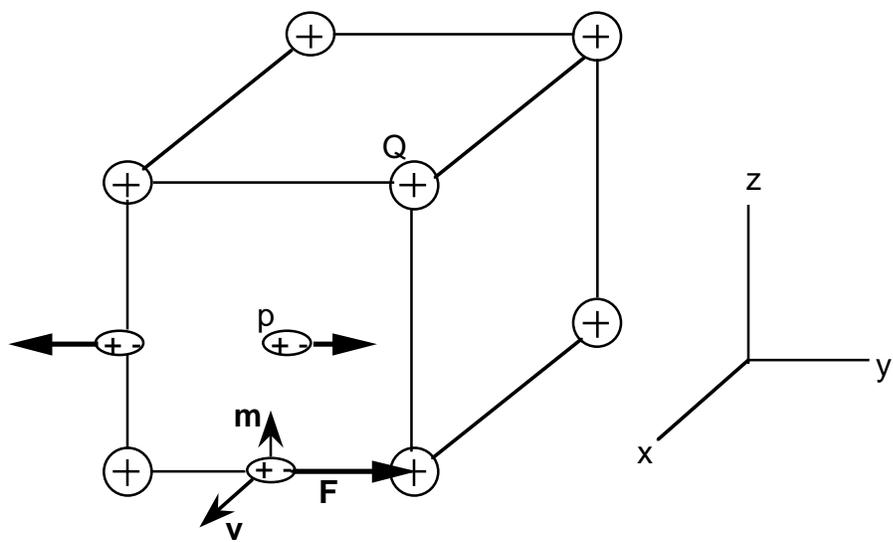
\begin{figure}
\caption {Simple cubic lattice of charges Q.  Magnetic moments pointing in
the z direction are propagating in the x direction and are equivalent to 
dipoles p shown. The direction of the force on the dipoles due to the periodic 
lattice of charges Q depends on the position of the dipoles
and is indicated by the bold arrows. The lenght of these arrows
indicates qualitatively the relative magnitudes for the three different
positions shown.
 }
\label{Fig. 2}
\end{figure}

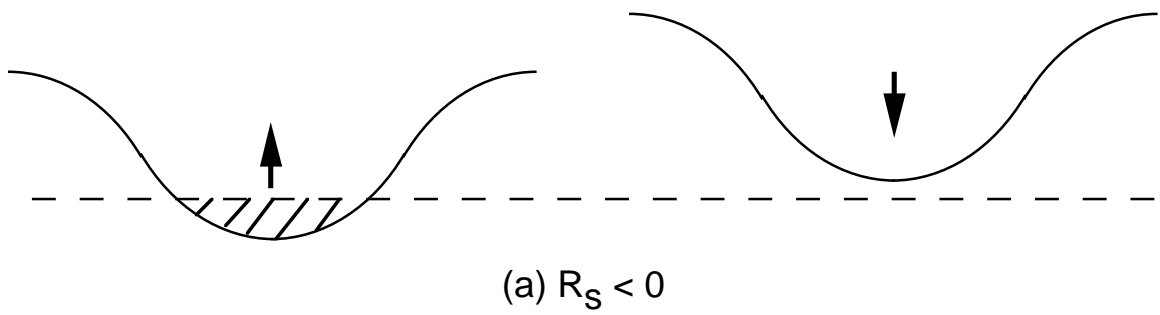
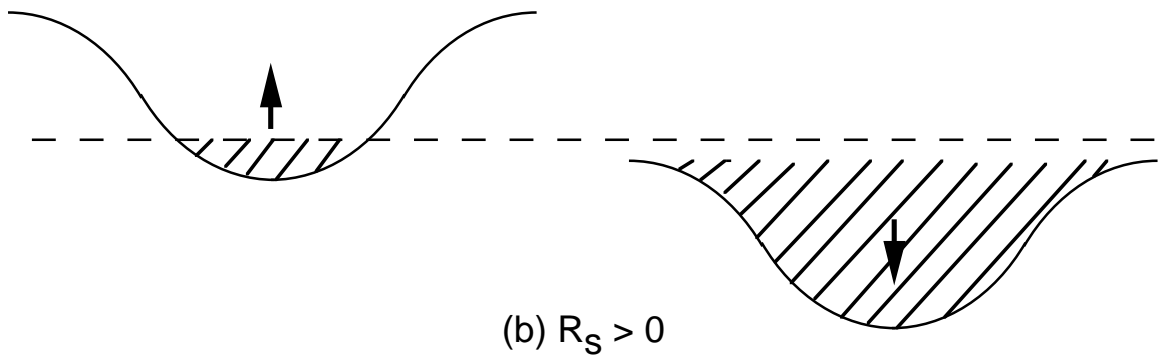
\begin{figure}
\caption {Examples where the anomalous and ordinary Hall coefficients
have the same sign (a) and opposite sign (b). In both cases $R_o<0$.
The dashed lines indicate the positions of the Fermi level.
In (a), the current carriers have magnetic moment parallel, in (b) antiparallel,
to the total magnetization. }
\label{Fig. 3}
\end{figure}

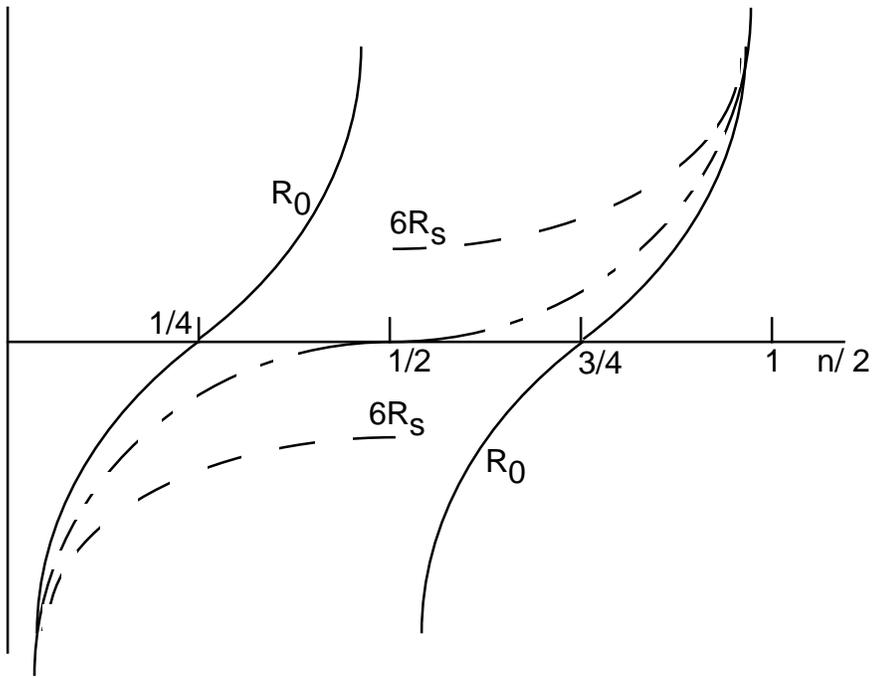
\begin{figure}
\caption {Ordinary  (full line) and anomalous (dashed line)  Hall coefficients
versus total band filling $n=n_\uparrow+n_\downarrow$
for a fully polarized single band symmetric around
its center (schematic). For unpolarized band filling less than one quarter 
($n<1/2$) and
more than three quarters ($n>3/2$) the ordinary and anomalous Hall coefficients
have the same sign. 
The long-dash short-dash line indicates the ordinary Hall coefficient
at onset of spin polarization.}
\label{Fig. 4}
\end{figure}


\begin{references}
\bibitem{explain} The dipoles themselves will also give rise to a transverse
electric field; however for the sample geometry used in the experiment, with
demagnetizing factor nearly unity, they will only give rise to a small transverse
voltage originating in the edge magnetic fields generated by the sample magnetization,
which does not increase with the sample width.
\bibitem{find} This result was found numerically to $9$ decimal places for
many values of $r$ and $\theta$. We have not been able so far to 
prove it analytically.
\end{references}
\end{document}